\documentclass{easychair}
\usepackage{cite}
\usepackage{float}
\usepackage{amsmath,amssymb,amsfonts}
\usepackage{soul}
\usepackage[linesnumbered,ruled,vlined]{algorithm2e}
\usepackage{graphicx}
\usepackage{textcomp}
\def\BibTeX{{\rm B\kern-.05em{\sc i\kern-.025em b}\kern-.08em
    T\kern-.1667em\lower.7ex\hbox{E}\kern-.125emX}}
\usepackage{hyperref}

\usepackage{doc}

\title{Security and Fairness in Multi-Party Quantum Secret Sharing Protocol}

\author{
Alessio Di Santo
\and
    Walter Tiberti
\and
   Dajana Cassioli
}

\institute{
  Università degli Studi dell'Aquila,
  L'Aquila, Abruzzo, Italy\\
  \email{alessio.disanto@graduate.univaq.it;walter.tiberti@univaq.it;dajana.cassioli@univaq.it}
}

\authorrunning{Di Santo, Tiberti and Cassioli}

\titlerunning{}

\begin{document}

\maketitle

\begin{abstract}
\textit{Quantum secret sharing} (QSS) is a cryptographic protocol that leverages quantum mechanics to distribute a secret among multiple parties. With respect to the classical counterpart, in QSS the secret is encoded into quantum states and shared by a \textit{dealer} such that only an authorized subsets of participants, i.e., the \textit{players}, can reconstruct it. Several state-of-the-art studies aim to transpose classical Secret Sharing into the quantum realm, while maintaining their reliance on traditional network topologies (e.g., star, ring, fully-connected) and require that all the $n$ \textit{players} calculate the secret. These studies exploit the Greenberger-Horne-Zeilinger (\textit{GHZ}) state, which is a type of maximally entangled quantum state involving three or more qubits. However, none of these works account for redundancy, enhanced security/privacy features or authentication mechanisms able to fingerprint players.
To address these gaps, in this paper we introduce a new concept of QSS which leans on a generic distributed quantum-network, based on a threshold scheme, where all the \textit{players} collaborate also to the routing of quantum information among them. The \textit{dealer}, by exploiting a custom flexible weighting system, takes advantage of a newly defined quantum Dijkstra algorithm to select the most suitable subset of $t$ \textit{players}, out of the entire set on $n$ players, to involve in the computation. To fingerprint and authenticate users, \textit{CRYSTAL-Kyber} primitives are adopted, while also protecting each \textit{player}’s privacy by hiding their identities. We show the effectiveness and performance of the proposed protocol by testing it against the main classical and quantum attacks, thereby improving the state-of-the-art security measures.
\end{abstract}

\section{Introduction}
\label{sec:introduction}
Secret sharing encompasses methodologies for distributing a secret among a group of individuals, each of whom does not possess any comprehensible information about it. Only when a requisite number of participants combine their respective shares, the original secret can be reconstructed. In contrast to insecure secret sharing, where an attacker can incrementally acquire more information with each share, \textit{secure secret sharing} adheres to an ‘all or nothing’ principle, where ‘all’ denotes the necessary number of shares needed to reconstruct the secret. Secret reconstruction can follow two distinct schemes. The first is the $(n,n)$ scheme, which relies on full participation, meaning every participant holds a piece of the secret, and it can only be recovered when all pieces are combined. The second is the $(t,n)$ scheme, which introduces a threshold, $t$, allowing a subset of participants (of size $t$) to reconstruct the secret without the involvement of the entire group.
Major studies on secret sharing began, in 1979, by G. R. Blakley and Adi Shamir. Blakley's research employs hyperplane geometry to address the secret sharing problem. To create a $(t,n)$ threshold scheme, each of the \textit{n} participants (2 or more) are provided with a hyperplane equation within a $t$-dimensional space over a finite field \cite{Blakley}. Shamir’s scheme, on the other hand, relies on a similar concept but is based on polynomial interpolation \cite{Shamir79}.

Since 1997, when Peter Shor demonstrated how modern mathematical pillars about hard-to-solve problems could be easily defeated by moving into the quantum realm\cite{Shor_1997}, researchers have felt the need to find additional security measures in the same domain that defeated them, i.e., the quantum one. Hence, the number of studies on \textit{Quantum Secret Sharing} \cite{Lu_2019, Hiesmayr_2021} and \textit{Quantum Fairness} \cite{Quantum_Comp, Liu_2014} grew since then. Nowadays, these kinds of studies focus on improving security and efficiency of their solutions, while addressing their vulnerabilities.

In 1999, a first class of QSS-protocols where explored, by exploiting \textit{Greenberger–Horne–Zeil-}\textit{inger} states \cite{Hillery_1999}. GHZ states are multipartite entangled states where multiple quantum particles exhibit strong, non-local correlations. Additional mathematical details and properties are provided in Appendix \ref{app:GHZ}

In recent years, Quantum Secret Sharing (QSS) protocols have gained significant attention for their potential in secure communications. However, despite advances, many existing protocols still face critical limitations that undermine their practicality in dynamic and high-security environments. These include rigid network topologies, do not taking into account for the need of a flexible routing mechanism, and no specific attention to \textit{players}' privacy. Furthermore, authentication mechanisms are basic and limited to participation phases, and fairness is only addressed in specific scenarios \cite{FS, Liu_2014, app10010189}. 

These shortcomings are extensively described in Section \ref{sota} and addressed by the proposed protocol, a \textit{dealer}-\textit{players} scenario based on a generic \textit{Entanglement-based Quantum Secret Sharing Protocol} over an adaptive quantum network architecture, through the implementation of a newly defined \textit{Quantum-Dijkstra} algorithm (correlated with a custom \textit{weighting system}), the usage of \textit{QKD}, \textit{CRYSTALS-Kyber}, and equipping it with the properties of \textit{fairness} and the \textit{CIA Triad (Extended)}.

\subsection{Paper's Contribution}
The core contributions of our research can be summarized as follows: 
\begin{itemize}
    \item Introduction of dynamic and flexible network topologies modeled on distributed computation.
    \item Enhanced player privacy by restricting network topology knowledge to reduce collusion.
    \item Development of a Quantum-Dijkstra algorithm for optimal participant selection.
    \item Establishment of a custom weighting system, which relies on both classical and quantum parameters, to support Quantum-Dijkstra's usage.
    \item Integration of CRYSTALS-Kyber for continuous post-quantum player authentication.
    \item Extension of fairness into a \textit{(t,n)}-scheme for tamper detection.
    \item Implementation of the extended CIA Triad Framework for a comprehensive information security.
\end{itemize}

The rest of the paper is organized as follows. In Section \ref{sota} a comparison with the state-of-the-art is proposed; in Section \ref{Prim} the proposed system model is shown; in Section \ref{protocol} proposed protocol is presented; in Section \ref{results} all the acquired results are discussed and analyzed; in Section \ref{addional} some additional considerations on quantum  and post-quantum cryptography are provided; in Section \ref{conclusion} final conclusions are presented. At the end, the Appendix will provide additional information on the most interesting mathematical concept discussed in this work.

\section{Related work} \label{sota}

According to the literature, two are the main methods to implement a QSS protocol: either via \textit{Entangled States} or via \textit{Mutually Unbiased (orthonormal) Bases}. Additional information on how these two core methods work and how do they relate are provided in Appendix \ref{app:EBMUB}. 

In this section both QSS schemes will be referenced to address what actually is the state-of-the-art and how it would be improved by our work.

C. Lu \textit{et al.} \cite{Lu_2019} propose a verifiable framework using entanglement-free states to build $(t,n)$ QSS schemes, aiming to overcome some limitations of conventional QSS schemes without entanglement, such as security vulnerabilities and ineffective cheating detection. Their solution involves encoding $k-1$ secrets using $k-1$ single quantum states and an additional quantum state for verification. By adding a verification quantum state, the framework can effectively thwart attacks such as wrong component embedding or eavesdropping, ensuring the accuracy of secret recovery by participants. 

S. Schauer \textit{et al.} \cite{Schauer_2010} proposed a variant of the standard HBB protocol by employing a \textit{dealer}, Alice, who randomly prepares a standard GHZ state and distributes the qubits to Bob and Charlie. An additional Z-basis is introduced for measurements. Bob and Charlie announce their bases and reveal some results to Alice for testing inequalities. Alice, without revealing the initial state, uses inequalities to detect eavesdroppers. The adjustments simplify the protocol by removing the need to check message order and improve efficiency with the introduction of the second GHZ state despite the additional Z-basis measurement, which can be minimized probabilistically. 

Joy D. \textit{et al}\cite{Joy_2019} discuss implementing the quantum secret sharing scheme introduced by Hillery \textit{et al.} \cite{Hillery_1999} using the five-qubit transmon bowtie chip ('ibmqx4'). They compare the experimental density matrix with the theoretical one via quantum state tomography and calculate the fidelity measure to assess the accuracy of the results. 

Choi M. \textit{et al.}\cite{Choi_2018} propose an $(n-1,n-1)$-threshold QSS protocol using an n qubit state approximating the GHZ state. In this protocol, $n$ players measure their qubits in the X or Y basis, similar to the HBB QSS protocol.

Even if these four studies have all introduced interesting and fascinating advancement in QSS studies, they preferred to orient their effort over static environment condition. However, addressing of how a protocol would behave under real-world quantum noise and dynamic conditions would be an interesting new point of view to identify both additional limitations and strengths of a proposed solution.

F. Liu \textit{et al.} \cite{Liu_2014} proposed a $(n,n)$-QSS protocol with \textit{fairness} where, given $k$ secrets, a $L$-bit check vector and $n$ \textit{players}, a \textit{dealer} Alice prepares $k(n+L)$ entangled states to be shared. \textit{Players} collaborates together by sharing one by one their share. At the end, as per the \textit{fairness} properties, either every one or no one gets the final result, with a probability of $25\%$ of having a cheater to disclose the master secret. While this work addresses fairness in QSS in a commending fashion, it would be of a scientific interest try to stretch out the protocol's execution conditions to gather how \textit{fairness} could be useful in a dynamically changing distributed network, where its topology and \textit{players} can continuously change.

Priyanka, V. \textit{et al.} \cite{Priyanka_QSS} built a QSS $(m,n)$-scheme which involves a dealer, a trusted reconstructor, and participants. The reconstructor manages secret recovery by aggregating shares contributed by participants (exploiting GHZ states), followed by a mutual verification process between the reconstructor and participants (using Quantum Fourier Transform). During secret recovery, participants integrate their information directly into the quantum system, ensuring secure data transfer and preventing theft. The introduction of the \textit{QFT} routine to identify participating \textit{players} is captivating, since it is not usual for this aspect to be considered in QSS scheme. However, this concept could still be enhanced by authenticating \textit{players} with a fixed, known and authenticated (possible by a \textit{Certification Authority}) key, rather than having \textit{players} calculating their auth-key from the acquired shared shard by applying the \textit{QFT} routine.

X. Li \textit{et al.} \cite{Li_2022} propose a protocol which takes advantage of a \textit{dealer} and a \textit{trusted third party} to shares secret shards with \textit{players}. The \textit{trusted third party} builds Entangled GHZ particles, verifies \textit{players} identities, and sends them the aforementioned states. Once their respective calculations are done, each \textit{player} sends the particle back to \textit{trusted third party} which will then measure them and validate the result with the \textit{dealer}. To authenticate \textit{players} this protocol uses an effective methodology based on the generation of a \textit{security identity number k}. Since no standard formulation on how to compute this are provided, an implementation could implement any desired function. Nevertheless, employing a \textit{Post-Quantum Cryptography} scheme to achieve such an authentication could provide an intriguing addition for QSS models.

L. Li \textit{et al} \cite{Li2024} propose an authenticated dynamic quantum multi-secret sharing scheme, where multiple secrets are packaged into a master secret using the Chinese Remainder Theorem and shared via a monotone span program. The scheme ensures authentication through quantum digital signatures based on entanglement swapping and supports dynamic updating of participants. It employs Pauli and Hadamard operators for encoding, providing robust security against various attacks. By employing a \textit{Quantum Digital Signature Schema}, this work provides a elegant improvement to QSS. At the same time, by considering how quantum information exchange is, nowadays, not very reliable for data transmission and may require several retransmissions to correctly deliver its content, it could be of interest to try to reduce the usage of Quantum communications to the bare minimum, as an example falling back to the usage of a \textit{Post-Quantum Cryptography} implementation.

Qin H. \textit{et al.}\cite{Qin2015} propose a quantum secret sharing (QSS) scheme that utilizes phase shift operations and Lagrange interpolation to realize a $(t,n)$ threshold structure. The dealer encodes the quantum state using a phase shift operation and sends it to the participants. Each participant then performs their phase shift operations based on their private keys, allowing any t out of n participants to reconstruct the original quantum state. This proposed methodology appears as striking and different from the previously discussed works, which implies as a pre-requisite the implementation of a \textit{Quantum Secure Direct Connection}. To let this protocol to better fit in a real-environment scenario, additional authentication considerations should be achieved, as also stated for some previous state-of.the-art-work by introducing a \textit{PQC} algorithm.

All the previously proposed works focus their attention on standard network topologies, without taking into account for dynamically changing environment and without exploring how to let a \textit{dealer}, when a third-party entity is considered for the protocol execution, to easily manage the identification of the subset of participating \textit{players}. \textit{Players} privacy, \textit{fairness} (except for \cite{Liu_2014}) and the establishment of the \textit{CIA Triad (Extended)} properties for protocols is also were not really considered a major concerning topic.

Our work was designed to take care of all the possible improvements that were highlighted in this section. Our protocol has been designed to be flexible and adapt to a dynamically variable distributed network, instead of relying on standard ones. It also employs additional security mechanism, empowers \textit{players} privacy, takes into account for retransmission in case of missing quantum information and provides the properties of \textit{fairness} and the \textit{CIA Triad (Extended)}. All of these were achieved with the aim to better let the protocol to blend in a real-world scenario. Furthermore, since the \textit{dealer} needs to manage the \textit{secrets} exchange between possibly different \textit{players} at each iterations, our proposed QSS scheme employs a newly defined \textit{Quantum-Dijkstra} algorithm, which exploits a custom \textit{weighting system}, to easily identify the most suitable subset of participating \textit{players}. At the same time, the weighting system was designed to be flexible and allow the dealer to dynamically adapt to network conditions (e.g., favoring directly-connected \textit{players} with respect to the ones requiring \textit{quantum-swaps} to be reached). To let the \textit{players} to join the network and eventually being drafted to the protocol, without relying on quantum identification algorithms or over a-priori secured means, they would need to reach the \textit{Certification Authority} that will employ the \textit{CRYSTALS-Kyber} Key Encapsulation Mechanism to provide the \textit{player} with a \textit{shared key}. This will now act as a \textit{Public key} that allows for authentication during the protocol.
\section{System model} \label{Prim}

\begin{figure}[t]
    \centering
    \includegraphics[width=0.6\linewidth]{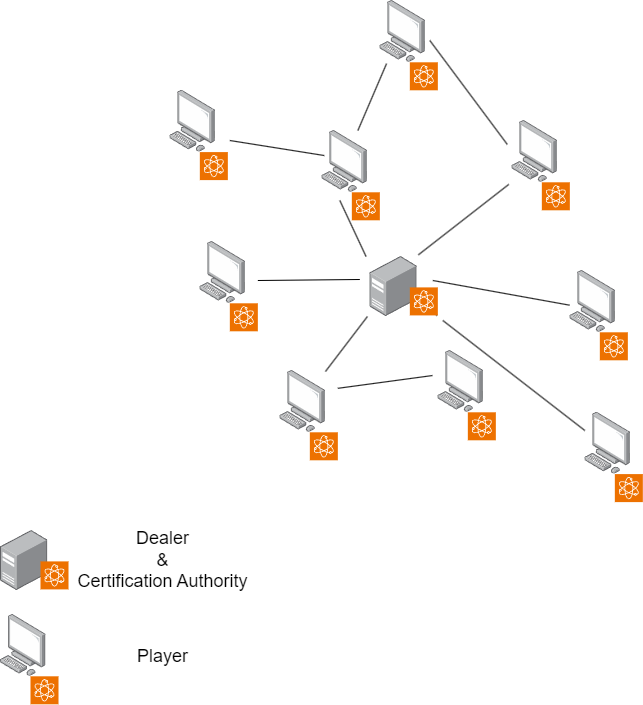}
    \caption{A Quantum-Network Topology for Secret Sharing.}
    \label{fig:network}
\end{figure}

We consider a Quantum-Network which involves $(n-1)$ \textit{players}, and a single \textit{dealer}, where all of them are equipped with quantum devices connected via optical fibers in the network topology shown in \figurename~\ref{fig:network}.

The \textit{dealer} acts as a \textit{Certification Authority} for the entire network and all participants are required to register themselves to be authenticated.
In the context of quantum networks, employing a distributed network topology rather than traditional structures like ring, common bus, or star configurations offers substantial advantages. Distributed topologies are characterized by nodes that relay information through a series of intermediate nodes rather than directly communicating with a central server.

In a distributed network, each node (or player) only needs to establish a connection with a single other node to become part of the network. This approach significantly enhances flexibility and scalability. When a new player joins the network, they simply connect to one existing player, and this new connection integrates this node into the network's communication fabric. This method avoids the bottleneck and single point of failure issues associated with centralized topologies, where every node must communicate directly with a central server.

In quantum networks, where the secure distribution of quantum information is paramount, relying on hops to reach a server through a distributed network can be more practical. This topology reduces the amount of direct communications needed between each node and the central server, thereby minimizing the potential for congestion and improving overall efficiency. Additionally, a distributed network allows for dynamic reconfiguration as nodes join or leave, making it more adaptable to real-world conditions compared to static topologies. This adaptability is crucial in quantum networks, where the integration of new participants and the maintenance of secure, reliable communication channels are essential.

\section{Quantum Secret Sharing Protocol} \label{protocol}

Inspired by \cite{Lu_2019, Priyanka_QSS}, the proposed protocol allows the \textit{dealer} to provide partial-secrets to a subset of $t$ players out of $n$, where one of them will be able to correctly reconstruct the secret by exploiting the \textit{Symmetric Polynomials} and \textit{Generalized Pauli Operators} \ref{app:EBMUB}.

To guarantee a $(t,n)$-threshold, the \textit{dealer} first generates a symmetric polynomial given by: 
\begin{equation}
    \begin{aligned}
        \mathbf{G}(x,y) &= \sum_{i=0}^{t-1} \sum_{j=0}^{t-1} a_{ij}x^i y^j \\
        &= a_{00} + a_{10}x + a_{01}y + \dots + a_{t-1,t-1}x^{t-1}y^{t-1}
    \end{aligned}
\end{equation}

with $\mathbf{G}(x,y) \in \mathbf{Z}_d$, where $d \in \mathcal{Z}$ | $d$ mod $2 \neq 0 \wedge d$ is prime, $deg(\mathbf{Z}_d) = t-1$ and $a_{00}$ is the \textit{secret} to reconstruct.

Next, the \textit{dealer} identifies the the most suitable $t$-dimensional subset of \textit{players} to involve in the protocol by executing the proposed Quantum-Dijkstra Algorithm described in Sec.~\ref{sec:QDJ}.

At this point, the dealer and selected players execute the proposed mutual authentication protocol described in Sec.~\ref{sec:Auth}
    
    Once this authentication phase is correctly achieved, any single \textit{player} is provided a \textit{secret} key to communicate with the \textit{dealer} (encrypting messages with a secure classical symmetric cipher as e.g., AES-256). From now on, every single message exchanged between \textit{dealer} and \textit{players} will always be encrypted with the corresponding \textit{secret} key.
    
    The \textit{dealer} will then share with the $i$-th player $P_i$ the polynomial$\mathbf{G}(x_i,y)$, with $i \in \{1, \dots, t\}$ and $x_i \in \mathcal{Z}_d$ being a random coefficient. This polynomial will later be needed to let the player to be able to reconstruct its \textit{share's shadow} and mathematically defined in the following lines.

The \textit{Entangled States Sharing} is done by the \textit{dealer} by building a $d$-dimensional (i.e., the local Hilbert space is isomorphic to $\mathbb{C}^{d}$) GHZ state in the form:
    \begin{equation}\label{eq:ghz}
        |\phi\rangle = \frac{1}{\sqrt{d}}\ \sum_{i=1}^{d-1} |\nu\rangle_1 \otimes \dots \otimes |\nu\rangle_{t-1}
    \end{equation}
    $|\phi\rangle$ contains exactly $t$ particles, one for each participating \textit{player}. The \textit{dealer} will indeed send $|\nu\rangle_i$ to all \textit{player} $P_i\, \forall\, i \in \{0,\dots t-1\}$. At this point, to avoid an attacker to be able to intercept the shared particle and disturb the protocol execution, $|\nu\rangle_i$ is not going to be shared alone in a single stream, although, additional arbitrary $j-1$ decoy particles $|\delta\rangle$ will be added to generate a $j$-particles stream, through a BB84-like protocol, to hide the real particle and avoid/decrease attacker's probability to gather any further information or invalidate this step of the protocol\cite{Bennett_2014, TN}. These steps are better explained in the following section:
:
    \begin{enumerate}
        \item The \textit{dealer}, while trying to transmit entangled particle $|\nu_i$ to \textit{player} $P_i$, generates $j-1$ random quantum decoy particles $|\delta\rangle$ and randomly polarize them in basis: $\{|0\rangle, |1\rangle, |+\rangle, |-\rangle\}$.
        \item \textit{Dealer} prepares the particle stream by placing, in a random spot, the real GHZ state $|\nu\rangle_i$ and obtaining a stream like the following: $|\delta\rangle_0 |\delta\rangle_1 \dots |\nu\rangle_i \dots |\delta\rangle_{j-1}$
        \item \textit{Dealer} builds a classical message explaining, for each particle, the measurement base and expected outcome, while also specifying where the entangled is placed to allow the \textit{player} to store it without measuring it, and sends it.
        \item Then, the \textit{dealer} begins by sharing one by one the particles;
        \item \textit{Player} applies the pre-shared measurements and verifies the expected match. The protocol will also take into account for a slightly probability of having one or more erroneous measurements due to swap actions and particle decoherence over the fiber channel. This value can be dynamically adapted to increase as the swaps do, to consider both security and protocol usability.
    \end{enumerate}
 
 Now, each participant $P_i$ can calculate its \textit{shares' shadows} $S_i$, i.e., the individual pieces of information distributed to players that collectively allow the reconstruction of the original secret, as follows:

    \begin{equation}\label{eq:shareshadow}
        S_i = \mathbf{G}(x_i,0) \prod_{j \neq i}^{t} \frac{x_j}{x_j - x_i} mod \, d
    \end{equation}
    
    All players compute the respective $|l_i\oplus S_i\rangle$ by embedding their $S_i$ equation, containing the secret, inside the previously received GHZ state $|\phi\rangle$, which gets modified as
    
    \begin{equation}\label{eq:phi1}
        \begin{aligned}
            |\phi_1\rangle &= \frac{1}{\sqrt{d}} \sum_{\nu = 0}^{d-1} d^{-\frac{t}{2}} 
            \sum_{l_1, \dots, l_t} \omega^{k(l_1,\dots,l_t)} \\
            &\quad \times (|l_1 \oplus S_1\rangle) \dots (|l_t \oplus S_t\rangle)
        \end{aligned}
    \end{equation}

    This can be achieved by using a QFT (Quantum Fourier Transform) circuit while exploiting the \textit{Generalized Pauli Operators} (further mathematical details provided in Appendix \ref{app:EBMUB}).

The players now measure the new obtained entangled particle in the computational base $\{0, \dots, d-1\}$ and obtain a measure $M_i = S_1 \oplus l_1$ each, ready to be shared with the other players.
When each single \textit{player} has embedded its \textit{secret} inside the shared entangled state, they need a new way to exchange the data and calculate the final value.

This can be done in two ways:
    \begin{itemize}
        \item \textit{Dealer acts as a distributor}: each single \textit{player} sends back to the \textit{dealer} its result by encrypting it with the pre-shared \textit{ss} key.
        The dealer deciphers and reads the measurements and verifies it is not forged. Otherwise, the protocol stops and, if a \textit{Fairness with penalties} configuration is used, cheating player might be fined.

        If no mismatches are detected, the dealer equally distributes all the shattered values to all the players as well as the hash of the secret, to let them to be able to confront protocol's result with it and confirm its legitimacy.
        
        \item \textit{Bulletin Board}: this idea was firstly exposed by \cite{PB}: it can be thought as an immutable object where only the dealer is authorized to post on and the players are only able to read. Hence, in this scenario the dealer's actions end after sharing $\mathbf{G}(x_i, y)$ polynomial and $|\nu_i\rangle$ GHZ particle, thus the dealer acts as a passive recipient whose main aim is just to publish the result. Hence, each \textit{player} sends its value to the \textit{dealer} which will push over a bulletin board the various results and the hash of the expected result, once everyone has sent its measurement.
    \end{itemize}

    Finally, the final \textit{secret} is reconstructed by each player as:
    
    \begin{equation}\label{eq:final}
        \begin{aligned}
            \sum_{i=0}^{t} M_i &= M_1 \oplus \dots \oplus M_t \\
            &= (S_1 \oplus l_1) \otimes \dots \otimes (S_t \oplus l_t) \\
            &= (S_1 \oplus \dots \oplus S_t) \otimes (l_1 \oplus \dots \oplus l_t) \\
            &= \sum_{i=0}^{t} S_i = S
        \end{aligned}
    \end{equation}
    
    Moreover, at the end of the protocol the hash of the secret (i.e., via SHA3) is provided to all the \textit{players}, to let them to check the correctness of the execution.

\subsection{Quantum Dijkstra Algorithm} \label{sec:QDJ}
\textit{Dijkstra} is a well-known path finding algorithm, and with this paper we reformulated it into a quantum version. Our main strategy is to leverage the Optimized Quantum Minimum Search Algorithm (OQMSA) \cite{Liu_2021} to introduce a new method for calculating the minimum value from a random vector. To provide a real enhancement to the classical Dijkstra algorithm we had to find where a quantum algorithm could speed-up the extraction of the best costing path and this was achieved by introducing \textit{OQMSA} as the new routine which extracts the minimum from the Dijkstra's paths list.

The Quantum Dijkstra algorithm is presented in Algorithm~\ref{alg:qdijkstra} and the OQMSA is described in Algorithm~\ref{alg:oqmsa}.

The introduction of the OQMSA approach, based on an enhanced and more accurate version of Grover's algorithm \cite{Long_2001}, reduces the time-complexity to $O(|V|*\sqrt{|V|})$ versus the $O(|V|^2)$ of the conventional Dijkstra's algorithm \cite{Dijkstra1959}. 
However, this efficiency gain is balanced against a small probability of extracting the incorrect minimum, because unlike classical algorithms, OQMSA can extract the exact minimum from an unsorted array with a success rate of 98\%.  

\begin{algorithm}[h]
\caption{Quantum-Dijkstra’s Algorithm with Edge Cost}\label{alg:qdijkstra}
\SetKwInOut{Input}{Input}
\SetKwInOut{Output}{Output}
\KwIn{Graph, Source}
\KwOut{dist, prev}

$vertices \gets \text{dim(Graph.Vertices)}$\;
\ForEach{$v \in Graph.Vertices$}{
    $dist[v] \gets |1\rangle^{\otimes vertices}$\;
    $prev[v] \gets null$\;
}

\While{$dim(Q) \neq |0\rangle$}{
    $u \gets OQMSA(dist[u])$\;
    $Q.pop(u)$\;
    \ForEach{$v \in Neighbors(Graph, u)$}{
        $C_{u,v} \gets \text{Graph.Edges}(u, v)$
        $|var\rangle \gets dist[u] + C_{u,v}$
        \If{$|var\rangle < dist[v]$}{
            $dist[v] \gets |var\rangle$\;
            $prev[v] \gets u$\;
        }
    }
}
\Return{$dist, prev$}\;
\end{algorithm}

    \begin{algorithm}[h]
    \caption{OQMSA \cite{Liu_2021}}\label{alg:oqmsa}
    \SetKwInOut{Input}{Input}
    \SetKwInOut{Output}{Output}
    \KwIn{$D$, $d'$ (Database and a Random Item)}
    \KwOut{$d_{min}$}
    
    \For{$i \gets 0$ \KwTo $\lceil \log(N) \rceil$}{
        $t_{max} \gets \left[\frac{\pi}{2} - \arcsin\left(\frac{1}{\sqrt{N}}\right)\right] / \arcsin\left(\frac{1}{\sqrt{N}}\right)$\;
        $t \gets 1$\;
        $\lambda \gets \frac{6}{5}$\;
        $r \gets \infty$\;
    }
    
    \If{$M/N > \frac{1}{9}$}{
        $t' \gets randint(0, \lceil t \rceil)$\;
        $|\phi'\rangle \gets Grover\_Long(|\phi\rangle, t')$\;
        $t \gets t * \lambda$\;
    }
    
    \If{$M/N < \frac{1}{9}$}{
        $|\phi'\rangle \gets Grover\_Long(|\phi\rangle, t_{max})$\;
    }
    
    $r \gets Measure(|\phi'\rangle)$\;
    
    \If{$r < d'$}{
        $d' \gets r$\;
        $i \gets 0$\;
    }
    
    \Return{$d_{min} = d'$}\;
    
    \end{algorithm}

This quantum Dijkstra algorithm uses specifically designed links' weights given by: 

\begin{equation} \label{formula}
    C_{v,u} = (\kappa/\alpha) + ((1-\kappa)*\beta))
\end{equation}

 where $v$ is a node in a  Graph $G$ and $u \in N$ $N = \mbox{Neighbors}(v, G)$.

In eq.~(\ref{formula}), $\alpha \in [0,1]$ represents a measure of the \textit{quality of entanglement}, i.e., it describes the probability of having a successful entanglement between the two ends of the connection. To let a lower entanglement swap success probability, i.e., $\alpha \to 0$, to increase the link's weight, the aforementioned formula will use the inverse of $\alpha$, i.e., $\frac{1}{\alpha}$. In such a way, for $\alpha \to 0$ we have $C_{v,u} \to \infty$.Since qubits cannot be copied, due to the No-Cloning Theorem, quantum swaps will need to occur to let \textit{players} to route entangled particles \cite{Quantum_Comp}. To model how $\alpha$ can be calculated it is possible to refer to \cite{ji2022, gitiaux2021}. 

The parameter $\beta$ in eq.~(\ref{formula}) accounts for the main characteristics of the fiber channel, i.e., capacity, non-linear interference noise, polarization-mode dispersion, and polarization-dependent loss \cite{Fibre}, and is calculated as shown in the pseudo-code snippet in Algorithm~\ref{alg:lookup_algorithm}.

\begin{algorithm}[h]
\caption{Channel Measurement and Lookup Table Matching}\label{alg:lookup_algorithm}
\KwIn{Nodes $u$, $v$}
\KwOut{Final value $\beta$}

\SetKwFunction{MeasureChannel}{MeasureChannel}
\SetKwFunction{LookupTable}{LookupTable}

$\beta \gets 0$\;
$\{P_1, P_2, \dots, P_n\} \gets List\, of\, channel\, parameters$

\ForEach{parameter $P_i$ in $\{P_1, P_2, \dots, P_n\}$}{
    $value_i \gets$ \MeasureChannel{$P_i$, $u$, $v$}
    
    $lookup\_value \gets$ \LookupTable{$P_i$, $value_i$}
    
    $\beta \gets \beta + lookup\_value$
}
\Return $\beta$\;
\end{algorithm}

Finally, the proposed formula in (\ref{formula}) involves the additional variable $\kappa$ to introduce another degree of flexibility. It will act as a weight to balance between a formula oriented on swap capabilities (i.e., Distributed Computing) over fiber's quality and vice-versa. 

If $\kappa \to 0$, then the process does not care about swapping qubits. Otherwise, if $\kappa \to 1$, it is needed to pass through links that allow for the best swapping results, without taking strictly into account all fiber's characteristics.

Equation (\ref{formula}) will then be employed to calculate each path cost, inside the proposed network, and build a \textit{Dijkstra} path-matrix which will allow the \textit{dealer} to identify the most suitable \textit{players} to involve in the protocol execution. Path cost computation can be achieved in two different ways, both of them are equally feasible inside this protocol and the protocol's settings will specify which one to use:
\begin{enumerate}
    \item \textit{Single Source Computation}: only the source node initiates the calculation of route costs. It iteratively updates the cost to reach each node based on the distances to neighbors. Once a node's minimum cost is determined, it is considered "visited," and the algorithm continues to the next unvisited node with the smallest cost.

    \item \textit{Distributed Variants}: In some distributed implementations every node may independently compute the cost to reach its neighbors. This can be useful in dynamic networks where topology changes frequently. Each node can then share its cost information with its neighbors, allowing for a more collaborative approach to computing paths but require additional \textit{player}-to-\textit{player} communications and hardware for \textit{players} quantum devices.
\end{enumerate}

\begin{algorithm}[h]
\caption{Weight Calculation Using Swap Success and Lookup Parameters}
\SetKwInOut{Input}{Input}
\SetKwInOut{Output}{Output}
\KwIn{$v, u, \kappa$}
\KwOut{$C_{v,u}$}

\SetKwFunction{EstimateSwapSuccess}{EstimateSwapSuccess}
\SetKwFunction{ParametersEstimation}{ParametersEstimation}

$\alpha \gets$ \EstimateSwapSuccess{$v, u$} \;

$\beta \gets$ \ParametersEstimation{$v, u$} \;

$C_{v,u} \gets \left( \frac{\kappa}{\alpha} \right) + \left( (1 - \kappa) * \beta \right)$\;

\Return{$C_{v,u}$}\;

\end{algorithm}

 The primary advantage of such a dynamical weighting system lies in its ability to dynamically adjust the selection of participants based on real-time network conditions, such as quantum noise levels, distance between nodes, and the reliability of individual quantum channels. This flexibility enables the system to prioritize more stable or efficient routes, ensuring that the quantum keys are securely and efficiently distributed, even in fluctuating network environments. 

An example of how the \textit{Decoding Error Probability}, over a quantum channel $\Lambda$, impacts over the computation of the $\beta$ parameter is provided in Table~\ref{tab:shape-functions}. Basing on the error level, $\beta$ will be added with corresponding \textit{score}.

Table~\ref{tab:shape-functions} takes into account also for the \textit{Quantum Information sent per qubit}, i.e., $\frac{1}{N}Q^{\epsilon}(\Lambda)$, which expresses the rate at which quantum information can be sent reliably through a quantum channel $\Lambda$ when the number of uses of the channel grows large, corresponding to a specific \textit{Decoding Error Probability} $\epsilon$.  

\begin{table}[b]
\centering
\begin{tabular}{cccc}
\hline
$N$ & Decoding Error probability ($\epsilon$) & $\frac{1}{N}Q^{\epsilon}(\Lambda)$ & Score\\
\hline
$10^5$ & $10^{-2}$ & $\approx 0.320$ & 3 \\
$10^5$ & $10^{-6}$ & $\approx 0.330$ & 4\\
$10^5$ & $10^{-10}$ & $\approx 0.360$ & 5\\
\hline
\end{tabular}
\centering
\caption{\bf Quantum information sent per qubit ($\frac{1}{N}Q^{\epsilon}(\Lambda)$)}
  \label{tab:shape-functions}
\end{table}

\begin{figure}[t]
    \centering
    \includegraphics[width=0.7\linewidth]{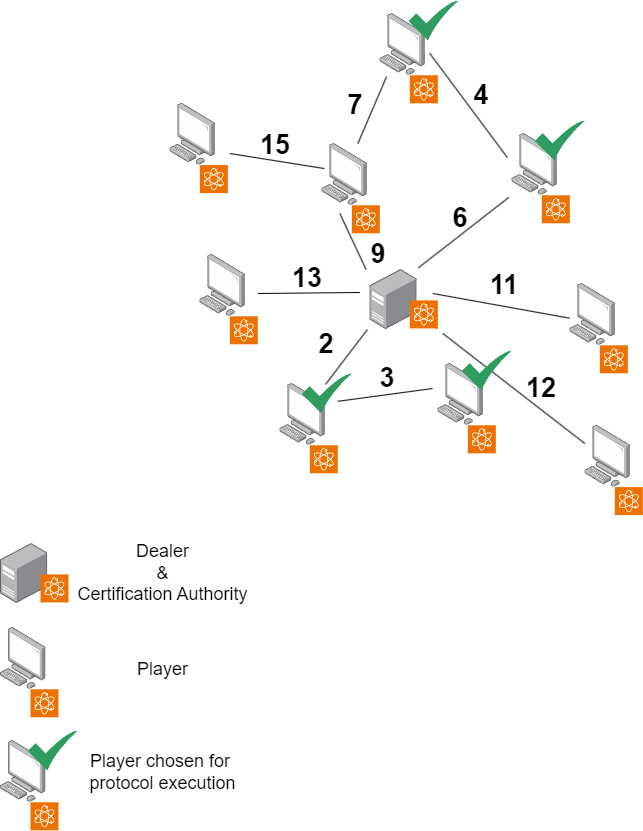}
    \caption{Distributed Network Topology with weighted links.}
    \label{fig:networkwork}
\end{figure}

Once the weighting system is applied to the entire network, a situation like the one proposed in \figurename~\ref{fig:networkwork} is gathered. At this point, the most suitable $t$ \textit{players} (4 in this example) are chosen to be participating to the next protocol iteration.

\subsection{Authentication} \label{sec:Auth}
\begin{figure}[t]
    \centering
    \includegraphics[width=0.9\linewidth]{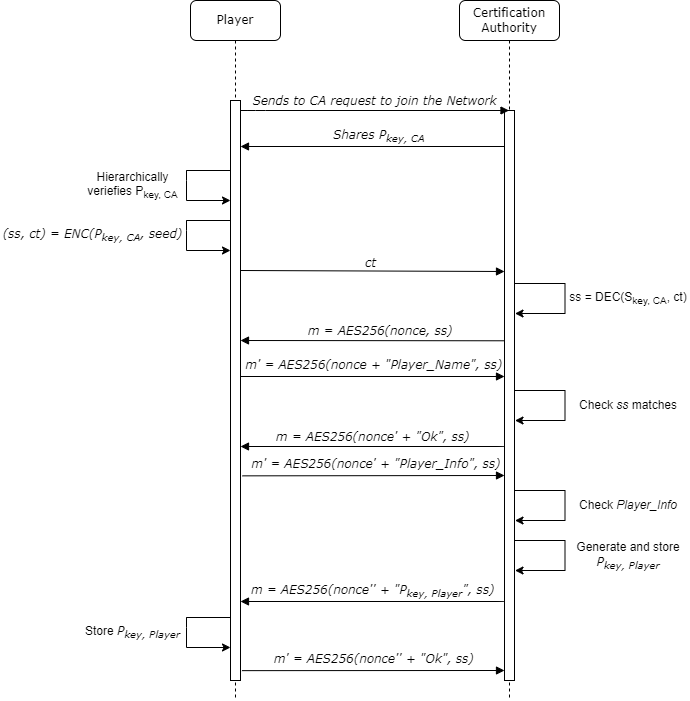}
    \caption{How a new Player joins the protocol.}
    \label{fig:kyber}
\end{figure}

For a new player to be admitted into the network, it must provide its \textit{Public Kyber Key} to the \textit{Certification Authority} (the \textit{dealer} in this scenario). The \textit{CA} will then register the new player. This process has been detailed in \figurename~\ref{fig:kyber}. 

\noindent

\begin{enumerate}
    \item A \textit{player} wants to join the protocol network and, to be accepted, it must at first register its public key with a \textit{Certification Authority}. By taking advantage of \textit{CRYSTALS-Kyber} primitives \cite{8406610}, \textit{CA} shares its \textit{Public Key} with the recipient;
    
    \item A \textit{participant} uses a hierarchically superior \textit{CA} to assess $P_{key, CA}$ authenticity;
    
    \item \textit{Participant} computes \textit{ciphertext} (ct) and a \textit{secret-shard} $ss$ by employing $P_{key, CA}$ and a random \textit{seed}. \textit{ct} is sent back to \textit{CA} to let it to compute \textit{ss} as well;
    
    \item To confirm that both parties agreed on the shared key, and to avoid any possible reply attack, \textit{Certification Authority} uses a symmetric encryption algorithm (i.e., AES-256) to encrypt a \textit{nonce} and send it to the \textit{Player}. If the latter has correctly executed the calculations, it will be able to decrypt the \textit{nonce} and send it back by attaching to it its name;
    
    \item \textit{CA} checks the $ss$ was correctly shared between them and sends back an "Ok" message to the recipient, meaning it is waiting for its information to begin with the key creation. The \textit{Player} does exactly what the \textit{Certification Authority} was waiting for and shares its personal information;
    \item \textit{CA} validates data and creates a new entry for this new acquired participant. Then, it randomly generates $P_{key, Player}$ and shares this key with the recipient.
    The \textit{Player}'s Public key is combined with a nonce and shared and an encrypted message. By achieving so, an eavesdropper is not able to fingerprint the key, avoiding to leaking information on who might get involved in the protocol.
\end{enumerate}

\section{Results} \label{results}
This section provides a comprehensive evaluation of the fairness mechanisms, and the security analysis of the Quantum Secret Sharing (QSS) protocol. 

Ensuring that no participant can gain an unfair advantage, the fairness aspect represents a critical requirement for secure multiparty computations. The results demonstrate how our protocol adheres to fairness by distributing quantum keys securely across all players, regardless of their position within the network.

Moreover, we explore the CIA Triad (Confidentiality, Integrity, Availability) within the protocol's framework. Our results confirm that the QSS protocol guarantees confidentiality through entanglement and encryption techniques, integrity by authenticating participants, and availability by maintaining system resilience in the face of network or quantum channel failures.

Finally, a security analysis evaluates the protocol against classical and quantum attacks, validating its robustness. This analysis provides evidence that the protocol not only achieves the necessary cryptographic strength but also advances beyond the existing solutions by integrating enhanced security, privacy, and authentication measures.

\subsection{Fairness}
In the proposed protocol, this constraint is achieved by relying on the followings:
\begin{enumerate}
    \item In the first protocol implementation, the \textit{dealer} acts as a \textit{broker} between all the parties. Indeed, it gets the measurements from them and, as first, checks for any forging attempts. Hence, all the \textit{players} are able to both gather the final result or know that someone tried to hijack it. For a cheater, it is difficult to correctly achieve its intent, since the \textit{dealer} knows which share has been given to it and which measurement is expected to be shared. In this case the role of this broker is dominant, but at the cost of a trusted third-party, not always available, there is a hidden security guarantee that can not be ignored;
    \item In the second implementation, which exploits the usage of a \textit{Bulletin Board}, the \textit{dealer} has a passive role, since it just acquires the measurements and then post them over a \textit{bulletin board}. Since the broker does not actively verify the correctness of the protocol, there could be a non-zero chance of the cheater being able to let all the fair players to gather wrong results.

    \noindent
    \textit{Dealer}'s final result hash is shared at the end not as a mere case, but it is chosen to avoid any possible brute-force attempts or to give any possible insight to the malicious \textit{player}. Nonetheless, the attacker(s) could still be able to generate an hash collision by providing a fake measurements:

    \noindent
    By considering how our proposed protocol works, and with the hypothesis of having $n$ total \textit{players}, $1$ \textit{dealer}, $t$ actively involved \textit{players}, let's assume there are $f$ fraudulent \textit{players}, with $t-f$ non-cheating \textit{players}. Cheaters could provide $M'_j, j \in \{0, \dots, f-1\}$ fake measurements with the objective of obtaining an hash collision, as shown in Equation \ref{eq:cheatobj}:.
    \noindent
    \begin{equation}\label{eq:cheatobj}
        \begin{aligned}
            \sum_{i=0}^{t} M_i &= S \\
            \Rightarrow \sum_{i=0}^{t-1-f} M_i &+ \sum_{j=0}^{f-1} M'_j = S' \\
            H(S) &= H(S')
        \end{aligned}
    \end{equation}

    At this point, if cheaters are able to correctly counterfeit the hash calculation, they will be the only one gathering the original \textit{secret}, while the remaining host parties will be provided with a fraudulent one.
\end{enumerate}

\subsection{CIA triad (extended)}\label{CIA}
The following paragraphs elucidates how each attribute of the Extended CIA Triad is demonstrated within the protocol, highlighting its comprehensive approach to safeguarding quantum communications.

\vspace{0.5em}
\noindent \textbf{Confidentiality}
The adoption of the CRYSTALS-Kyber Key Encapsulation Mechanism (KEM) allows both the Dealer and players to mutually authenticate each other and securely establish a shared encryption key. This key, generated during the registration phase, serves as a unique identifier for each participant, thereby enhancing the security of the system. Once established, this shared key is used to ensure data confidentiality through a symmetric encryption scheme, such as AES-256. Since Kyber is a post-quantum algorithm designed for classical environments, its integration with symmetric encryption like AES-256 is more straightforward compared to quantum-based cryptographic schemes, which typically require complex setups and may be more vulnerable to noise and implementation challenges.

A different analysis needs to be made for GHZ states. Even if an attacker is able to capture and read its value (destroying the particle) it will not gather any useful information, since the secret calculation is made by the \textit{player} once it received the entangled particle. Hence, any manipulation can only cause a delay in the protocol's execution;
All the classical information exchanged between the parties can rely over a, e.g., generic TCP-fashion network communication where each datagram is equipped with a CRC-like code to allow for integrity checks. 

Instead, while dealing with quantum particles exchange, a different mechanism needs to be applied to improve the integrity of the shared entangled state. To do so, in the proposed scheme the \textit{dealer} adds \textit{decoy particles} while sharing GHZ states with \textit{players}. This allows to estimate the likelihood of an attacker to hijack the protocol. The more particles are incorrectly received by the \textit{player}, the more probabilities there are that the communication is under attack. If too many particles do not align with the expected outcome, the entire process is repeated, allowing for managing the integrity of the delivered $|\nu\rangle_i$ state.

\vspace{0.5em} \noindent \textbf{Availability}.
In classical network communication, information could, when needed, re-transmitted. Quantum particle distribution, on the other hand, is more susceptible to volumetric attacks (e.g., Denial-of-Service attacks). However, if at least one of the \textit{players} has not received its the entangled particle the \textit{dealer} will ask every \textit{player} to stop and restart the \textit{Entangled States Sharing} phase. The protocol performance will be degraded but it would be still operative and able to be correctly completed;

\vspace{0.5em} \noindent \textbf{Authenticity}.
As per the previously mentioned authentication mechanism, based on a \textit{Certification Authority} and the \textit{CRYSTALS-Kyber} KEM, each \textit{player} authenticates the \textit{dealer} and the latter authenticates all the involved \textit{players};

\vspace{0.5em} \noindent \textbf{Accountability}.
Since each message signed with a specific \textit{secret share}, both \textit{dealer} and \textit{players} are able to confirm their identities, as per the usage of \textit{CRYSTALS-Kyber}. To also take into account for attacks where a \textit{player} is stolen of its keys, \textit{network probes} can be placed in the network intercepting traffic and allowing for further review of the packets' source, allowing for any rogue node to be identified;

\vspace{0.5em} \noindent \textbf{Non-repudiation}.
Once a message has been signed with a specific \textit{secret share}, it is not possible for a \textit{player} or even the \textit{dealer} to repudiate it. Since the \textit{ss} is owned by them (generated with the aid of a pair of public/private keys), no other can use it to sign messages (if no attacks involving key-starling activities are achieved).

\vspace{0.5em} \noindent \textbf{Security Analysis}.
This section aims to provide a detailed examination of the protocol's security features, offering insights into its strengths:
\begin{itemize}
    \item \textit{Denial of Service}: in the proposed scheme, an attacker may attempt to disrupt either quantum or classical communications, or both. The quantum network under consideration functions as a distributed system, with multiple hosts not necessarily directly connected to the \textit{dealer}. This design introduces an additional layer of redundancy; even if an attacker targets one or more classical or quantum channels, the \textit{dealer} can respond by removing \textit{players} under DoS attack, in favor of incorporating new ones. Moreover, the network topology is not publicly disclosed, and each player is aware only of their immediate neighbors. Consequently, rogue entities can only impact a limited portion of the network. Additionally, as discussed in relation to the protocol's \textit{accountability} feature, strategic placement of \textit{network probes} allows for traffic monitoring and the potential to \textit{mute} non-collaborating nodes, thereby maintaining network integrity and security.
    
    \begin{figure}[t]
    \centering
    \includegraphics[width=0.7\linewidth]{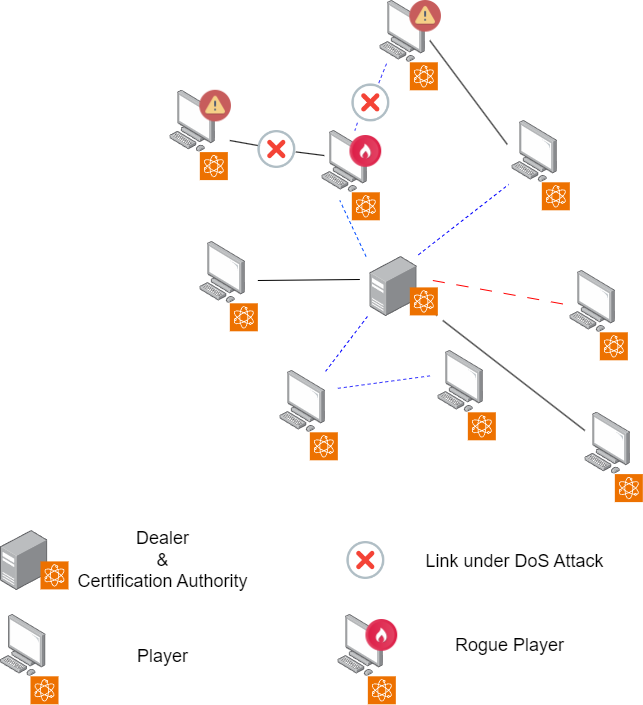}
    \caption{How a DoS can impact proposed network.}
    \label{fig:DoS}
    \end{figure}

    As illustrated previously, \textit{players} participating in a protocol round are indicated with a blue dotted link. A rogue \textit{player} may attempt a Denial of Service (DoS) attack on nearby connections, obstructing the \textit{dealer} from reaching them. However, if a player is compromised, the protocol adapts by selecting another player based on the \textit{Quantum-Dijkstra} algorithm, indicated by a red dotted link. This approach ensures the protocol's resilience and continuity despite potential disruptions.
    \noindent
    Generally, quantum communications are inherently more susceptible to interference. An attacker could attempt to delay the protocol by estimating which of the $v$ quantum-exchanged particles contains the GHZ state. By identifying and replacing the genuine particle with a fraudulent one, a rogue node can disrupt the protocol, causing it to fail due to the loss of entanglement necessary for embedding the \textit{secret shards} within a random photon. Probabilistically, an attacker altering a single particle has, at most, a $\frac{1}{v}$ chance of successfully targeting the entangled particle without affecting the control particles. Moreover, the attacker might estimate a higher likelihood of the GHZ state being placed towards the end of the stream due to the photon's unstable nature and susceptibility to rapid decoherence. Thus, as $v$ increases, so does the likelihood of the entangled state being located in the latter part of the sequence. To mitigate this, $v$ should be carefully chosen to securely obscure the entangled particle while not being too large to avoid facilitating tailored guessing by the attacker.
    
    \item \textit{Reply attack}: To prevent a malicious entity from reusing pre-exchanged information to deceive another party several countermeasures are employed. During the registration of a new \textit{player} and protocol's AUTHENTICATION phase, \textit{nonces} are used. Consequently, an attacker cannot rely on previously shared messages, as each message is based on a\textit{nonce} and a different \textit{shared secret} (ss), which is re-generated for each protocol run. Additionally, private keys are never publicly shared, ensuring that no attacker can recover them without access to a participant's device. Therefore, an attacker has only two potential methods to introduce themselves into the protocol:
    
    \begin{itemize}
        \item \textit{Entangle shared particles}: \textit{Dealer} sends a \textit{player} $v$ photons. Each of them can be intercepted by an attacker, locally entangled, and then let to flow to the recipient. In this way, the rogue node disposes of something it can use to try to gather further information. Anyway, since the encoding information and the GHZ state position are shared as an encrypted message, this fraudulent actor does not know which particle is to be measured and which one to be used as the embedding source. Hence, it gathers no further useful insights;
        \item \textit{Intercept and Resend Particles}: An attacker might try to measure all $v$ photons and then re-encode the information, performing a \textit{man-in-the-middle} attack. However, the security constraints are analogous to those of the \textit{BB84} protocol. Without prior knowledge of the encoding basis, the attacker has a $25\%$ chance of guessing the correct measurement basis for each particle. For a stream of $v-1$ particles, where the entangled particle must not be measured, the probability of correctly guessing all measurement bases is $\frac{1}{4}^{(v-1)}$, making the attack highly improbable.
    \end{itemize}
    
    \item \textit{Spoofing}: to compromise the protocol by spoofing a participant's identity, an attacker would need to obtain a \textit{private key} or break the \textit{CRYSTALS-Kyber} primitives. Currently, no known methods exist to easily solve the lattice problems underlying this encryption scheme, making such an attack highly impractical.
    
    \item \textit{Collusion}: Given $n$ \textit{players}, $1$ \textit{dealer}, $t$ participating hosts, $f$ fraudulent players (with $f < t$) providing $M'_j, j \in \{0, \dots, f-1\}$ fake measurements they are able to fix the result iff:

    \noindent
    \begin{equation}
        \begin{aligned}
            \sum_{i=0}^{t} M_i &= S \\
            \Rightarrow \sum_{i=0}^{t-1-f} M_i &+ \sum_{j=0}^{f-1} M'_j = S' \\
            H(S) &= H(S')
        \end{aligned}
    \end{equation}

    Hence, a sub-group of rogue \textit{players} can threaten the protocol if, and only if, they are able to create an hash-collision with their fake measurements. In this case, only the malicious \textit{players} would know the real protocol's output.

    \item \textit{Trojan Horse}: Given that the proposed protocol introduces a BB84-like photon exchange to conceal the entangled particle and enhance rogue node detection, it is essential to consider the potential for a \textit{Trojan-Horse} attack scenario. This issue becomes even more significant when there are one or more relay nodes between the \textit{dealer} and the \textit{player}, as this impacts the protocol's security.

    \noindent
    Ideally, all light entering a receiving system would seamlessly transfer to the output via interfaces and components. However, in real-world scenarios, some light may inevitably be reflected or scattered back while passing through an interface. Fresnel reflections occur due to variations in the refractive index during propagation, while Rayleigh or Brillouin scattering results from density fluctuations in the optical fiber material. The wavelength and intensity of the incoming light affect the amount of scattering and reflection. An eavesdropper might introduce a light pulse via the quantum channel into a Quantum Key Distribution (QKD) subsystem, encountering multiple sites of reflection and scattering. This could allow the attacker to gather information on how the sender polarized the last emitted particle.

    \noindent
    While it might seem overzealous to discuss additional security measures given the QKD-like method used in the GHZ states' sharing phase of the protocol, it is important to note that even though the states' polarization does not carry any information, detecting attempts to mine the exchange is crucial. Since relays are part of the quantum-network definition, enhancing security also involves adding layers to detect unfair behavior.
    
    \noindent
    To increase security against \textit{Trojan-Horse} attacks, one approach is to minimize the emitter's \textit{opening-frame} time to reduce the eavesdropper's time-span. To block and detect an attack, a combination of an \textit{isolator} and a \textit{watchdog} can be used. This setup ensures that incoming pulses are either blocked or detected, alerting the \textit{players} and identifying the rogue node. Additionally, an optical filter (such as Bragg gratings (FBGs) or Fabry-Perot cavities) can be added after the monitoring detector. Physical constraints, such as using angle polished connectors (FC/APC) instead of flat connectors (FC/PC), can also reduce light back-scattering, further enhancing security against potential \textit{Trojan-Horse} attacks.

    \begin{figure}[t]
        \centering
        \includegraphics[width=0.75\linewidth]{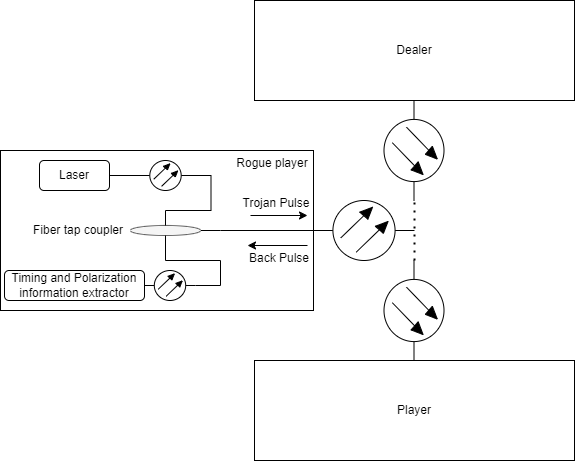}
        \caption{Trojan Horse attack implementation\cite{TH}.}
        \label{fig:THA}
    \end{figure}

\end{itemize}

\section{Open challenges} \label{addional}
\subsection{Post-Quantum Cryptography}
\textit{Learning with Errors} (\textit{LWE}) is a fundamental problem in cryptography that involves solving systems of linear equations, where small errors (or noise) are intentionally added to the equations. This noise makes the system much harder to solve than traditional systems of linear equations. In essence, \textit{LWE} turns solving these equations into a computationally difficult problem, even for quantum computers, which is why it forms the basis for many post-quantum schemes.

The relationship between \textit{LWE} and \textit{Kyber} lies in \textit{Kyber}'s security foundation. \textit{Kyber} is a lattice-based Post-Quantum Cryptography (PQC) algorithm, and its security is based on the hardness of solving LWE problems. Lattice-based cryptography is believed to be resistant to attacks from quantum computers, which is why \textit{Kyber}, relying on LWE, has been standardized by NIST as a secure key encapsulation mechanism (KEM) in a post-quantum world.

\textit{LWE}’s strength is that even quantum computers cannot efficiently solve this problem, making it an ideal foundation for encryption schemes designed to secure communications in the post-quantum era. Therefore, \textit{Kyber}’s reliance on LWE offers robust security against classical and quantum attacks.

However, it is crucial to consider whether contemporary researchers are facing a scenario similar to that of their counterparts in 1977 when \textit{RSA} was first introduced. This reflection is necessary to understand that the perceived quantum-proof nature of this cryptosystem is closely tied to the current absence of efficient quantum algorithms for solving LWE. There are no studies affirming the impossibility of breakthroughs against LWE. Therefore, the main concerns regarding asymmetric cryptosystems remain unresolved and merely postponed. It is conceivable that, eventually, lattice problems could be reformulated into a problem akin to "finding the period of a function." Should this occur, a variant of Shor's Algorithm might be employed to effectively compromise the cryptosystem in a very short time.

\subsection{Quantum Cryptography}
\textit{Quantum Cryptography} is widely considered as the next frontier and something researchers should dig into. Nowadays, there is a tendency into adapting milestones of classical computing to fit into quantum realm (as it has been done in this research with \textit{Quantum-Dijkstra}). Anyway, probably, studies should evolve and try to create new concepts not relegated to classical constraints. This means that, as an example, creating a quantum version of AES is interesting and fascinating, as done egregiously by \cite{designinghashencryptionengines}, but what would happen if the entire core behind a symmetric encryption scheme is not borrowed from classical cryptography and created as an entirely new quantum concept? A first sight of this dissertation is shown by BB84 and E91. Both of them do not have a classical counterpart, anyway they are still very advanced concept that were able to gather fame and admiration in the cryptographic field.

\noindent
Obviously, there is also a good reason behind a sense of caution against quantum-cryptography, and it is located in the insufficient confidence researchers got with it. Classical cryptographic schemes were studied $7^{th}$ century BC and analysts developed a widely broad set of primitives and mathematical concepts able to assess the security of a newly developed scheme. With quantum ones, all this prior knowledge does apply anymore, since new rules and constraints were placed. This means that, at the moment, governments and companies deal with this new technology as something to study but to not fully trust. Hence, further studies need to be taken and as soon as possible, since classical primitives got an expiration date and to protect security and privacy of everyone, aforementioned solutions will be needed to be applied as soon as possible.

\section{Conclusions} \label{conclusion}
By addressing several critical limitations of existing models, a new entanglement QSS protocol was proposed, enhancing the flexibility and resilience of quantum network topologies, which are often overlooked in static network models. By ensuring that players are unaware of both their collaborators and the network structure, the protocol effectively mitigates the risk of collusion and is thus suitable for deployment in highly secure environments.

Furthermore, the innovative use of the Quantum-Dijkstra algorithm for selecting the subset of $t$ players introduces an efficient and adaptable method for participant selection, taking into account various practical factors such as fiber propagation properties and proximity to the dealer.

The integration of the CRYSTALS-Kyber post-quantum cryptographic scheme for player authentication strengthens the protocol's security framework. This mechanism ensures that only authorized participants are involved, thereby enhancing trust and reducing the risk of identity spoofing.

Additionally, the emphasis on fairness within a (t,n)-scheme, rather than the traditional (n,n)-scheme, underscores the importance of verifying the correctness of results and detecting tampering attempts. This approach further secures the integrity of the protocol against potential cheating.

Lastly, by addressing the extended CIA Triad attributes—Confidentiality, Integrity, Availability, Authenticity, Accountability, and Non-repudiation—the proposed protocol provides a robust framework for comprehensive information security. This holistic approach ensures that all aspects of information exchange are protected, thereby fortifying the protocol against a wide range of security threats.

Overall, the proposed entanglement-based QSS protocol demonstrates a balanced and rigorous approach to enhancing both the security and practical applicability of quantum communication networks, paving the way for more secure and efficient quantum information sharing.

\noindent
As future works, provided concepts could be expanded by improving the Quantum-Network Information Exchange, the Algorithms used (e.g., Dijkstra weights calculation scheme) and also modeling a \textit{Semi-Definite Programming Model} able to describe the possibilities held by a one single party to be able to completely cheat the protocol.
\appendix
\section{Greenberger-Horne-Zeilinger (GHZ) state}\label{app:GHZ}
A Greenberger-Horne-Zeilinger (GHZ) state is a type of multi-particle entangled quantum state that plays a central role in quantum information theory, especially in studies of quantum entanglement and nonlocality. It is named after physicists Daniel Greenberger, Michael Horne, and Anton Zeilinger, who first formulated it as a generalization of Bell's theorem to more than two particles.

For three qubits (the simplest non-trivial GHZ state), the GHZ state is given by the following superposition:
\begin{equation} \label{eq:GHZ}
    \begin{aligned}
        |GHZ\rangle = \frac{1}{\sqrt{2}} (|000\rangle + |111\rangle)
    \end{aligned}
\end{equation}

Here, $|000\rangle$ represents the state where all three qubits are in the basis state $|0\rangle$, and 
$|111\rangle$ represents the state where all three qubits are in the basis state $|1\rangle$. In general, the n-qubit GHZ state is written as:

\begin{equation} \label{eq:GHZd}
    \begin{aligned}
        |GHZ\rangle = \frac{1}{\sqrt{2}} (|0\rangle^{\otimes n} + |1\rangle^{\otimes n})
    \end{aligned}
\end{equation}

Here, there will be reported all the main properties regarding GHZ states, most of them were actually used in our proposed protocol to allow the \textit{secret sharing} routine to be securely achieved:
\begin{itemize}
    \item \textit{Maximal Entanglement}: GHZ states exhibit maximal entanglement among all qubits. If any single qubit of the GHZ state is measured, the entanglement collapses, meaning that the remaining qubits become disentangled. The measurement outcome for one qubit will determine the state of all others.

    \item \textit{Nonlocality}: GHZ states show a strong form of quantum nonlocality. Unlike Bell's theorem, which shows nonlocal correlations for two entangled particles, GHZ states extend this concept to multiple particles, providing an even stronger conflict between quantum mechanics and classical local realism. The violation of local realism becomes even more pronounced as the number of qubits increases.

    \item \textit{Quantum Superposition}: The GHZ state is in a superposition of two distinct states, with no intermediate states. This means that if a measurement is made in the computational basis, the result will always be either all qubits in state |0⟩ or all qubits in state |1⟩—never a mix. This property of superposition leads to the nonlocal correlations observed in experiments.

    \item \textit{Fragility}: GHZ states are highly fragile in the presence of noise or decoherence. Any small disturbance can quickly destroy the entanglement between qubits. This sensitivity to environmental effects makes them challenging to maintain over long periods or in complex systems.

    \item \textit{Measurement Outcomes}: In a GHZ state, measurement in different bases reveals different types of quantum correlations. For example, if measurements are made in the computational basis (|0⟩, |1⟩), the outcomes will be highly correlated, but in other bases (such as the Pauli-X or Pauli-Y basis), more complex entanglement properties are revealed.

    \item \textit{Symmetry}: GHZ states exhibit symmetry with respect to the interchange of any of the qubits. This means that the entanglement is not localized between specific qubits, but rather, the entanglement involves the entire system equally.
\end{itemize}

\section{Quantum Secret Sharing core implementations}\label{app:EBMUB}
In the following section we compare the two core methods to implement a QSS protocol, i.e., \textit{Mutually Unbiased Bases} and \textit{Entangled States}. 
Both methods are useful to advance the security and efficacy of quantum secret sharing protocols, each addressing different aspects of quantum information security.

\textit{Entangled States}, discussed in \cite{Schauer_2010, Joy_2019, Choi_2018, Liu_2014, Priyanka_QSS, Li_2022, Schmid_2005, Gottesman_2000}, involve quantum particles that are correlated in such a way that the state of one particle is instantaneously linked to the state of another, regardless of where they are positioned in space. In QSS schemes, entanglement is used to distribute quantum secrets among multiple participants. For example, in entanglement-based QSS protocols like those inspired by E91 \cite{PhysRevLett.67.661}, entanglement enables the secure sharing of information by creating a situation where the quantum state can only be reconstructed by a specific number of participants. The correlations between entangled particles ensure that the information remains secure and any attempt to eavesdrop or tamper with the particles is detectable.

\textit{Mutually Unbiased Bases} are examples of \textit{Entanglement-free} schemas, discussed in \cite{Lu_2019, Li2024, Qin2015, Tavakoli_2015, Karimipour_2015, Semiquantum}. \textit{MUBs}, are sets of orthonormal bases in a quantum system where measurement outcomes in one basis provide no information about measurements in another. This property is leveraged in QSS schemes to enhance security, particularly in quantum key distribution. For instance, in protocols like BB84 \cite{Bennett_2014}, MUBs ensure that an eavesdropper cannot infer any information about the key being shared because the measurement results in one basis do not reveal any details about the results in another. This feature of MUBs ensures that the secrecy of the key is maintained even if an adversary attempts to intercept the quantum communication. To mathematically define this concept it is needed to refer to the following.

Given two orthonormal bases 
\begin{equation} \label{MUB}
    \mathcal{F} = \{|b_0\rangle, \dots, |b_d\rangle\} \\ \textnormal{ and } \\ \mathcal{C} = \{|c_0\rangle, \dots, |c_d\rangle\}
\end{equation} 
where $F,C \in \mathcal{H}^d$ (Hilbert space of dimension d) they are defined:

\noindent
\textit{Mutually Unbiased Basis} $\iff {|\langle b_i|c_i\rangle|}^2 = \frac{1}{\sqrt{d}}, \forall i,j \in {1, \dots, d}$.

Consider a finite-dimensional Hilbert space $\mathcal{H}$, s.t. $dim(\mathcal{H} = N)$, with $N < \infty$, and a \textit{Generalized Pauli Operators} by equipping the space $\mathcal{L}(\mathcal{H})$ with linear operators acting on $\mathcal{H}$ with two unitary operators as generators of the group: $\widehat{A}$ and $\widehat{B}$, where: 

\noindent
$\widehat{A}\widehat{B} = \omega^{-1}\widehat{B}\widehat{A}$, with $\omega = e^{\frac{2\pi i}{N}}$ is a primitive root of unity\cite{singh2020modeling}.

If $\mathcal{H}$ is a finite d-dimensional Hilbert space wit basis in $\{|0\rangle, \dots, |d-1\rangle\}$, the \textit{Generalized Pauli Operators} can be expressed as the following Unitary operator:

\noindent
\begin{equation}
    U_{a,b} = \sum_{y=0}^{d-1} \omega^{b,y}x^iy^j = |y \oplus a\rangle\langle y|
\end{equation}
where

\noindent
\begin{equation}
    U_{i,0}|y\rangle = |y\oplus i\rangle,U_{0,i}|y\rangle = \omega^{i,y}\langle y|
\end{equation}
For $i \in \{0, \dots, d-1\}$, $U_{i,0}$ is known as generalized Pauli X gate, while $U_{0,i}$ as generalized Pauli Z one.

The relationship between \textit{MUBs} and \textit{Entangled States}, in QSS, lies in their complementary roles in ensuring security. While MUBs are utilized to guarantee that measurements reveal no information about each other, thereby enhancing the security of key distribution, entangled states provide a means to securely share and reconstruct quantum secrets based on quantum correlations. The key difference between them is their approach to security: MUBs focus on measurement uncertainty to protect key information, while entangled states use quantum correlations to secure the distribution and reconstruction of secrets. Hence, entanglement provides stronger security guarantees, as the correlations are non-local and cannot be replicated without cooperation from all parties. Although managing entanglement requires careful handling to prevent degradation, the superior protection against eavesdropping and enhanced error correction capabilities make entanglement-based schemes more advantageous for high-security applications, despite their complexity.
\newpage
%\section*{Acknowledgment}
%This work is supported by the project ISP5G+ (CUP D33C22001300002), which is part of the SERICS program (PE00000014) under the NRRP MUR program funded by the EU-NGEU.

\bibliographystyle{unsrt}
\bibliography{biblo}

\begin{thebibliography}{10}

\bibitem{Blakley}
G.~R. Blakley.
\newblock Safeguarding cryptographic keys.
\newblock {\em 1979 International Workshop on Managing Requirements Knowledge
  (MARK)}, pages 313--318, 1979.

\bibitem{Shamir79}
A.~Shamir.
\newblock How to share a secret.
\newblock {\em Communications of the ACM}, 22(11):612--613, 1979.

\bibitem{Shor_1997}
P.~W. Shor.
\newblock Polynomial-time algorithms for prime factorization and discrete
  logarithms on a quantum computer.
\newblock {\em SIAM Journal on Computing}, 26(5):1484--1509, Oct. 1997.

\bibitem{Lu_2019}
C.~Lu, F.~Miao, J.~Hou, W.~Huang, and Y.~Xiong.
\newblock A verifiable framework of entanglement-free quantum secret sharing
  with information-theoretical security.
\newblock {\em Quantum Information Processing}, 19(1), 2019.

\bibitem{Hiesmayr_2021}
B.~C. Hiesmayr.
\newblock Free versus bound entanglement, a np-hard problem tackled by machine
  learning.
\newblock {\em Scientific Reports}, 11(1), 2021.

\bibitem{Quantum_Comp}
R.~van Houte, J.~Mulderij, T.~Attema, I.~Chiscop, and F.~Phillipson.
\newblock Mathematical formulation of quantum circuit design problems in
  networks of quantum computers.
\newblock {\em Quantum Information Processing}, 19(5), 2020.

\bibitem{Liu_2014}
F.~Liu, S.-J. Qin, and Q.-Y. Wen.
\newblock A quantum secret-sharing protocol with fairness.
\newblock {\em Physica Scripta}, 89(7):075104, 2014.

\bibitem{Hillery_1999}
M.~Hillery, V.~Bužek, and A.~Berthiaume.
\newblock Quantum secret sharing.
\newblock {\em Phys. Rev. A}, 59(3):1829--1834, 1999.

\bibitem{FS}
G.~Gao.
\newblock Cryptanalysis and improvement of efficient multiparty quantum secret
  sharing based on a novel structure and single qubits.
\newblock {\em EPJ Quantum Technol.}, 11, 2024.

\bibitem{app10010189}
Y.~Kang, Y.~Guo, H.~Zhong, G.~Chen, and X.~Jing.
\newblock Continuous variable quantum secret sharing with fairness.
\newblock {\em Appl. Sci.}, 10(1):189, 2020.

\bibitem{Schauer_2010}
S.~Schauer, M.~Huber, and B.~C. Hiesmayr.
\newblock Experimentally feasible security check for n-qubit quantum secret
  sharing.
\newblock {\em Phys. Rev. A}, 82(6), 2010.

\bibitem{Joy_2019}
D.~Joy, M.~Sabir, B.~K. Behera, and P.~K. Panigrahi.
\newblock Implementation of quantum secret sharing and quantum binary voting
  protocol in the ibm quantum computer.
\newblock {\em Quantum Inf. Process.}, 19(1), 2019.

\bibitem{Choi_2018}
M.~Choi, Y.~Lee, and S.~Lee.
\newblock Quantum secret sharing and mermin operator.
\newblock {\em Quantum Inf. Process.}, 17(10), 2018.

\bibitem{Priyanka_QSS}
P.~Priyanka, V.~Siwach, and P.~Bijaranian.
\newblock Quantum secret sharing with (m, n) threshold: Qft and identity
  authentication, 2024.

\bibitem{Li_2022}
X.~Li, K.~Zhang, L.~Zhang, and X.~Zhao.
\newblock A new quantum multiparty simultaneous identity authentication
  protocol with the classical third-party.
\newblock {\em Entropy}, 24(4):483, 2022.

\bibitem{Li2024}
L.~Li, Z.~Han, Z.~Li, F.~Guan, and L.~Zhang.
\newblock Authenticable dynamic quantum multi-secret sharing based on the
  chinese remainder theorem.
\newblock {\em Quantum Inf. Process.}, 23(2):46, 2024.

\bibitem{Qin2015}
H.~Qin, X.~Zhu, and Y.~Dai.
\newblock (t,n) threshold quantum secret sharing using the phase shift
  operation.
\newblock {\em Quantum Inf. Process.}, 14(8):2997--3004, 2015.

\bibitem{Bennett_2014}
C.~H. Bennett and G.~Brassard.
\newblock Quantum cryptography: Public key distribution and coin tossing.
\newblock {\em Theor. Comput. Sci.}, 560:7--11, 2014.

\bibitem{TN}
F.~Li, T.~Chen, and S.~Zhu.
\newblock A (t,n) threshold quantum secret sharing scheme with fairness.
\newblock {\em Int. J. Theor. Phys.}, 62(6):119, 2023.

\bibitem{PB}
A.~R. Choudhuri, M.~Green, A.~Jain, G.~Kaptchuk, and I.~Miers.
\newblock Fairness in an unfair world: Fair multiparty computation from public
  bulletin boards.
\newblock In {\em Proceedings of the 2017 ACM SIGSAC Conference on Computer and
  Communications Security}, pages 719--728, 2017.

\bibitem{Liu_2021}
W.~Liu, Q.~Wu, J.~Shen, J.~Zhao, M.~Zidan, and L.~Tong.
\newblock An optimized quantum minimum searching algorithm with sure-success
  probability and its experiment simulation with cirq.
\newblock {\em Journal of Ambient Intelligence and Humanized Computing},
  12(11):10425--10434, Jan. 2021.

\bibitem{Long_2001}
G.~L. Long.
\newblock Grover algorithm with zero theoretical failure rate.
\newblock {\em Phys. Rev. A}, 64(2), 2001.

\bibitem{Dijkstra1959}
E.~W. Dijkstra.
\newblock A note on two problems in connexion with graphs.
\newblock {\em Numerische Mathematik}, 1(1):269--271, Dec 1959.

\bibitem{ji2022}
Z.~Ji, P.~Fan, and H.~Zhang.
\newblock Entanglement swapping theory and beyond, 2022.

\bibitem{gitiaux2021}
X.~Gitiaux, I.~Morris, M.~Emelianenko, and M.~Tian.
\newblock Swap test for an arbitrary number of quantum states, 2021.

\bibitem{Fibre}
M.~Shtaif, C.~Antonelli, A.~Mecozzi, and X.~Chen.
\newblock Challenges in estimating the information capacity of the fiber-optic
  channel.
\newblock {\em Proceedings of the IEEE}, 110(11):1655--1678, 2022.

\bibitem{8406610}
J.~Bos, L.~Ducas, E.~Kiltz, T.~Lepoint, V.~Lyubashevsky, J.~M. Schanck,
  P.~Schwabe, G.~Seiler, and D.~Stehle.
\newblock Crystals - kyber: A cca-secure module-lattice-based kem.
\newblock In {\em 2018 IEEE European Symposium on Security and Privacy
  (EuroS\&P)}, pages 353--367, 2018.

\bibitem{TH}
Y.~Pan, L.~Zhang, and D.~Huang.
\newblock Practical security bounds against trojan horse attacks in
  continuous-variable quantum key distribution.
\newblock {\em Appl. Sci.}, 10(21):7788, 2020.

\bibitem{designinghashencryptionengines}
S.~Upadhyay, R.~Roy, and S.~Ghosh.
\newblock Designing hash and encryption engines using quantum computing.
\newblock 2023.

\bibitem{Schmid_2005}
C.~Schmid, P.~Trojek, M.~Bourennane, C.~Kurtsiefer, M.~Żukowski, and
  H.~Weinfurter.
\newblock Experimental single qubit quantum secret sharing.
\newblock {\em Phys. Rev. Lett.}, 95(23), 2005.

\bibitem{Gottesman_2000}
D.~Gottesman.
\newblock Theory of quantum secret sharing.
\newblock {\em Phys. Rev. A}, 61(4), Mar 2000.

\bibitem{PhysRevLett.67.661}
A.~K. Ekert.
\newblock Quantum cryptography based on bell's theorem.
\newblock {\em Phys. Rev. Lett.}, 67(6):661--663, Aug 1991.

\bibitem{Tavakoli_2015}
A.~Tavakoli, I.~Herbauts, M.~Żukowski, and M.~Bourennane.
\newblock Secret sharing with a single d-level quantum system.
\newblock {\em Phys. Rev. A}, 92(3), 2015.

\bibitem{Karimipour_2015}
V.~Karimipour and M.~Asoudeh.
\newblock Quantum secret sharing and random hopping: Using single states
  instead of entanglement.
\newblock {\em Phys. Rev. A}, 92(3), 2015.

\bibitem{Semiquantum}
F.~He, X.~Xin, C.~Li, and F.~Li.
\newblock Security analysis of the semi-quantum secret-sharing protocol of
  specific bits and its improvement.
\newblock {\em Quantum Inf. Process.}, 23, 2024.

\bibitem{singh2020modeling}
A.~Singh and S.~M. Carroll.
\newblock Modeling position and momentum in finite-dimensional hilbert spaces
  via generalized pauli operators.
\newblock 2020.

\end{thebibliography}

\end{document}